# Evolution of Nanoporosity in Dealloying

Jonah Erlebacher*†, Michael J. Aziz*, Alain Karma**, Nikolay Dimitrov***, Karl Sieradzki***

*Division of Engineering and Applied Sciences, Harvard University, 9 Oxford St., Cambridge, MA 02138, USA*

**Department of Physics and Center for Interdisciplinary Research on Complex Systems, Northeastern University, 360 Huntington Avenue, Boston, MA 02115, USA*

***Department of Mechanical and Aerospace Engineering and Center for Solid State Sciences, Arizona State University, Tempe, AZ 85287-6106, USA*

†Present address: Department of Materials Science and Engineering, Johns Hopkins University, Baltimore, MD 21218, USA.

**Correspondence and requests for materials should be addressed to Jonah Erlebacher (e-mail: Jonah.Erlebacher@jhu.edu.).**

Dealloying is a common corrosion process during which an alloy is "parted" by the selective dissolution of the electrochemically more active elements. This process results in the formation of a nanoporous sponge composed almost entirely of the more noble alloy constituents[1]. Even though this morphology evolution problem has attracted considerable attention, the physics responsible for porosity evolution have remained a mystery[2]. Here we show by experiment, lattice computer simulation, and a continuum model, that nanoporosity is due to an intrinsic dynamical pattern formation process - pores form because the more noble atoms are chemically driven to aggregate into two-dimensional clusters via a spinodal decomposition process at the solid-electrolyte interface. At the same time, the surface area continuously increases due to etching. Together, these processes evolve a characteristic length scale predicted by our continuum model. The applications potential of nanoporous metals is enormous. For instance, the high surface area of nanoporous gold made by dealloying Ag-Au can be chemically tailored, making it suitable for sensor applications, particularly in biomaterials contexts.

Selective dissolution has a long and rich history[3]. The chemical treatment known as depletion gilding, for instance, selectively dissolves a non-gold element near the surface of a less expensive alloy such as Au-Cu, leaving behind a surface of pure gold. Early Andean metalsmiths used this technique to enhance the surfaces of their artifacts[4]. In this century, selective dissolution has been primarily examined in the context of corrosion. It is observed in technologically important alloy systems, notably brasses, stainless steels, and Cu-Al alloys[1,5,6]. The mechanical properties of a porous overlayer are very different from the bulk alloy on which it sits, leading to brittle crack propagation, stress corrosion cracking, and other undesirable materials failure[7]. Figure 1 shows the prototypical dealloyed microstructure, that of nanoporous gold (NPG). Early notions considered porosity as a hidden microstructure revealed by etching but diffraction experiments showed that no pre-existing length scale exists prior to acid attack of single-phase alloys[8,9]. Later ideas considered the influence of percolating clusters within the solid solution of the alloy, but models failed to yield behavior consistent with experiment[10,11].

The following argument illustrates the fundamental roadblock to understanding porosity formation during dealloying: Consider a silver-gold alloy in an electrolyte under conditions where silver dissolves and gold is inert. Initially, silver will be dissolved from surface sites such as terraces or steps. Gold atoms should accumulate on the surface and locally block further dissolution. For a 10% gold alloy, it might be expected that dissolution would stop or be significantly retarded after about 10 monolayers of the alloy have been dissolved.

A complete model of selective dissolution is multi-scale, involving the kinetics of dissolution, surface diffusion, and mass transport through the bulk of both alloy and electrolyte. Because mass transport through the bulk of the growing phase (the electrolyte) is always a stabilizing influence[12] and mass transport through the bulk of the dissolving phase appears too slow to be significant, we hypothesized that the morphology-determining physics is confined to the interface region between the alloy and the electrolyte. To test this, we developed a kinetic Monte Carlo (KMC) model to simulate Ag-Au dealloying, including only diffusion of silver and gold and dissolution of silver[13]. Remarkably, we reproduced all relevant experimental trends characteristic of dealloying, both morphological and kinetic.

Figure 2 shows a simulated porous structure with 2-5 nm ligament widths. The simulations were successful in modeling the nanoporous morphology, and also in modeling the dynamic behavior of the dissolution current vs. overpotential. It is a well characterized feature of alloy dissolution that as the overpotential is ramped up (usually at rates of order mV/sec), the dissolution current of ions from the alloy stays at a low level until reaching a bulk composition-dependent critical potential $V_C$, at which point it rises rapidly[14]. Figures 3 shows simulated and experimental polarization curves for different alloy compositions. There is clear observation of a composition dependent $V_C$. This is the first simulation model to produce such behavior, suggesting that we have found a minimum set of physics to include in any model for alloy dissolution.

The simulations reveal the following qualitative picture of porosity formation. The process starts with the dissolution of a single silver atom on a flat alloy surface of closed pack (111) orientation, leaving behind a terrace vacancy. The atoms coordinating this vacancy have fewer lateral near neighbors than other silver atoms in the terrace, and are thus more susceptible to dissolution. As a result, the entire terrace is "stripped", leaving behind gold atoms with no lateral coordination ("adatoms"). Before the next layer is attacked, these gold adatoms diffuse about and start to agglomerate into islands. As a result, rather than a uniform diffuse layer of gold spread over the surface, the surface is comprised of two distinct kinds of regions, namely, pure gold clusters that locally passivate the surface, and patches of un-dealloyed material exposed to electrolyte. When silver atoms in the un-dealloyed patches dissolve into solution, more gold adatoms are released onto the surface. These diffuse to the gold clusters left over from dissolution of previous layers, continuing to leave un-dealloyed material exposed to electrolyte. In the early stages, these gold clusters are mounds that are gold rich at their peaks and that have alloy composition at their bases. These mounds get undercut, increasing the surface area that gold must cover to bring about passivation. Ultimately, this leads to pit formation and full-blown porosity. A movie of a simulation of the early stages of this process is available at http://www.deas.harvard.edu/matsci/downdata/downdata.html .

Central to this description is the observation of the coalescence of gold adatoms into stable clusters. The spacing between these "islands" in the initial stages of dissolution is close to the spacing between ligaments in the final porous structure. The physical reason for this coalescence can be understood by considering the gold adatoms to be one component of a two-component solution of gold and "electrolyte" confined to the monolayer-thick interfacial layer sitting on top of un-dealloyed material. We modeled the thermodynamics of the interfacial layer as a regular solution[15], and found that the solubility of gold in electrolyte within the interfacial layer is of order $10^{-7}$/site (see Methods). One may interpret this solubility as the "equilibrium concentration of gold adatoms" on the surface of the alloy – in the absence of etching, it represents a dynamic equilibrium of adatoms resulting from their two-dimensional evaporation from step edges onto terraces and their subsequent recondensation.

In contrast to the equilibrium condition, rapidly stripping a terrace of silver atoms leaves gold adatoms with a local site occupancy fraction equal to that in the bulk, typically 10-40% - far above their equilibrated concentration of $10^{-7}$/site. Thus, there is an extremely strong driving force for gold adatoms to condense onto nearby gold-rich clusters. In fact, regions of surface with high enough supersaturation of gold adatoms sit "within the spinodal," a special segment of the curve representing free energy $f$ of a spatially uniform layer vs. gold concentration $C$ for which $\partial^2 f/\partial C^2 < 0$. Within the spinodal, composition fluctuations of infinitesimal amplitude lead to a lower overall free energy for the system, and involve atomic diffusion against concentration gradients (the "uphill diffusion" process through which gold condenses onto nearby clusters), i.e., the system is inherently unstable and will spontaneously phase separate. Long length scale fluctuations, however, grow slowly due to the required diffusion times, and short length scale fluctuations create a lot of energetically unfavorable incipient interface between the phases, inhibiting their growth. Hence, phase separation is manifested most rapidly at an intermediate length scale that roughly corresponds to the spacing between the observed gold-rich clusters. This effect is known as spinodal decomposition[16,17]. As porosity forms, the decomposition is occurring on a non-flat, non-uniform surface with continuously increasing surface area.

The motion of the alloy-electrolyte interface is fully described mathematically by the flux of diffusing adatoms $J_S$, the velocity of the interface normal to itself $v_n$, and the concentration accumulation rate $\partial C/\partial t$, all of which are interrelated and vary along the arclength of the surface (for detailed derivations, see Methods). For $J_S$, we used a model for diffusion during spinodal decomposition known as the Cahn-Hilliard equation (CH) [17]. The normal velocity depends on $C$ and also on the local curvature $\kappa$ through capillary effects[11]. The time evolution of $C$ is uniquely determined by the local mass conservation condition

$$\partial C/\partial t = v_n C_0 - v_n \kappa C - \nabla \cdot J_S, \qquad (1)$$

where $C_0$ is the bulk gold concentration. This condition is analogous to the local conservation of heat or solute appearing in boundary-layer models of solidification[18], with two important differences: (i) the interfacial layer thickness is constant along the interface, and microscopic, versus being a spatially varying macroscopic diffusion length, and (ii) the surface aggregation process inherent in the CH form for $J_S$ is essential for porosity formation. Simple ("downhill") surface diffusion ($J_S = -D_S \nabla C$) yields an initially unstable interface that passivates quickly, before well-formed pores have a chance to develop. CH diffusion also dominates capillarity-driven surface diffusion - the effect usually incorporated into interface evolution equations[19].

We performed numerical integration of Eq. (1) using a relative arclength parameterization scheme[20], and parameters that matched those used in the KMC simulations. We observed, as expected, the evolution of gold clusters separated by a characteristic spacing $\lambda$. An analytic expression for $\lambda$ can be found by a time-dependent linear stability analysis of Eq. (1) that takes into account the slow increase of gold concentration into the interfacial layer as the instability develops. This effect needs to be included because the spatial period with the largest amplification rate depends sensitively on the gold concentration. Specifically, this spatial period decreases sharply as the concentration increases past a threshold concentration for instability that corresponds to the spinodal point $\partial^2 f/\partial C^2 = 0$, and the interface is stable for concentrations below this threshold. We find a maximally unstable spatial period that scales as $\lambda \propto (D_S/V_0)^{1/6}$, where $V_0$ is the velocity of a flat alloy surface with no gold accumulated upon it. This prediction is in qualitative agreement with both KMC simulations and experiments, both of which show that the characteristic length scale of porosity decreases with increasing driving force. A more elaborate analysis incorporating nonlinear effects, however, remains needed for a detailed quantitative comparison.

There is an interesting analogy between this result, applicable to etching, and two-dimensional island nucleation during submonolayer vapor phase deposition. Namely, in the early stages of etching, the dissolution process is analogous to deposition of gold; in both processes, adatoms are added to the surface where they are free to agglomerate into islands. The case of vapor phase deposition has been studied using rate equations that describe an aggregation process where adatoms stick together irreversibly. In these studies it is a well-known result that the island spacing scales as $(D_S/F)^\mu$, where the deposition rate $F$ is the direct analog of the surface velocity in etching, and the exponent $\mu$ depends on details of the aggregation process[21]. That these results are limited to irreversible aggregation during deposition and our analysis is for reversible aggregation during etching suggests the existence of universal scaling laws for aggregation that do not depend on reversibility or the lack thereof in these two opposite processes.

The KMC simulation elucidates the later stages of morphological evolution, and the mechanism by which 3D porosity evolves. We highlight the features of this process by showing in Figure 4 a simulation of an artificial pit in an otherwise fully passivated surface. When the pit reaches sufficient depth, its surface area has increased sufficiently that a new gold cluster nucleates. When this happened, the pore splits into multiple new pits, each with a smaller surface area than its parent. These "child" pits continue to penetrate into the bulk, increasing their surface area, nucleating new clusters, spawning new pits, etc., until a full 3D porous structure evolves such as those illustrated in Figures 1 and 2.

**Methods**

In a regular solution, the enthalpy of mixing depends on the bond energies and the entropy of mixing is ideal. The free energy of a regular solution $f(C,T)$ is written $f = \alpha c(1-c) + k_B T[c \ln c + (1-c)\ln(1-c)]$, where $c$ is the mole fraction of gold ($c = C\Omega^{2/3}$, $\Omega =$ atomic volume), $\alpha = 6(E_{Au\text{-}Electrolyte} - (1/2)(E_{Au\text{-}Au} + E_{Electrolyte\text{-}Electrolyte}))$, $E_{x\text{-}x}$ are the respective interaction energies between Au and electrolyte, the prefactor 6 is the lateral coordination in the 2D hexagonal lattice of the interfacial layer, $k_B$ is Boltzmann's constant and $T$ is absolute temperature. For our simulation conditions, realistic time- and length-scales were obtained from the parameters $E_{Au\text{-}Au} = -0.285$ eV ( $= -\varepsilon$, the simulation bond energy as described in Figure 2), $T = 600$ K, $E_{Au\text{-}Electrolyte} = E_{Electrolyte\text{-}Electrolyte} = 0.0$ eV. With these parameters, $\alpha = 0.855$ eV. The free energy has the familiar double-well form[24] and a minimum at $c \sim 10^{-7}$ site$^{-1}$, representing the solubility of gold in electrolyte (and vice versa).

The CH diffusion equation is $J_S = -M(C)(\partial^2 f/\partial c^2)\nabla C + 2M(C)w\nabla^3 C$. Here, $M(C)$ is a mobility, $w$ is the so-called gradient energy coefficient, and the gradients are taken with respect to arc length. The first term on the right hand side describes the chemical effect leading to phase separation within the spinodal; the second term describes the effect that damps short wavelength fluctuations. The mobility is proportional to the surface diffusivity $D_S$ and is given by $M(C) = (D_S/k_B T)c(1-c)$. The mobility is peaked for $c = 0.5$ and zero for $c = 0$ and $c = 1$ (atoms don't diffuse in pure phases because there are no vacancies in our model). The normal velocity is given by $v_n(C) = V(C)[1-(\gamma\Omega/k_B T)\kappa]$, where $\gamma$ is the surface free energy and $V(C)$ is called the interface response function, equal to the velocity of a flat surface covered with a concentration $C$ of gold. We find in both simulation and experiment that the interface response is fit well by the functional form $V(C) = V_0(\phi)\exp(-C/C^*)$, where $\phi$ is the overpotential and $C^*$ is a constant. Experimentally, one can infer the gold accumulation by integrating the dissolution current vs. time at fixed overpotential. In this approach one has be careful to be at sufficiently low overpotential that the surface remains planar (i.e., porosity does not form) and also to catch the short initial transient rise in current as silver atoms are pulled from the first few monolayers. This particular form for the interface response function is quite curious. Naively, one might expect that the local interface velocity would be proportional to the local concentration of silver exposed to the electrolyte, i.e., $V(C) \propto (1-c)$. However, the decaying exponential form suggests that there is an evolving distribution of holes opening and closing within the interfacial region, controlling the accumulation rate.

Physically, the mass conservation condition (Eq. 1) is the statement that the total number $Cb\Delta s$ of gold atoms in a length $\Delta s$ of interface with lateral width $b$ can change as a result of three distinct effects that correspond to the three terms on the right-hand-side of Eq. (1): the accumulation of gold atoms into the interfacial layer from the solid being dissolved; the local stretching of the interface ($\partial \Delta s/\partial t = v_n \kappa \Delta s$), which can either increase or decrease $C$ depending on whether the solid is concave ($\kappa > 0$) or convex ($\kappa < 0$); and the motion of atoms along the interface driven by the surface diffusion flux $J_S$.

## Acknowledgements

We thank the United States Department of Energy, Basic Energy Sciences for research support. The research of A.K. also benefited from computer time allocation at NU-ASCC.

Figure Captions

Figure 1. SEM micrographs of nanoporous gold made by selective dissolution of silver from Ag-Au alloys immersed in nitric acid under free corrosion. (A) cross-section of dealloyed $Au_{32\%}Ag_{68\%}$ (at %) thin film. (B) Plan-view of dealloyed $Au_{26\%}Ag_{74\%}$ (at %). The porosity is open and the ligament spacings shown in the micrograph Figure 1B are of order 10 nm; spacings as small as 5 nm have been observed. Measurements of the surface area of NPG are of order 2 $m^2/g$[22,23], comparable to commercial supported catalysts.

Figure 2. Simulated nanoporous gold. The simulation model is described as follows: A bond-breaking model was used for diffusion; atoms with $N$ near neighbors diffused with rate $k_N = \nu_D \exp(-N\varepsilon/k_B T)$, where $\varepsilon$ is a bond energy and $\nu_D = 10^{13}$ $sec^{-1}$. Dissolution rates were consistent with the Butler-Volmer (BV) equation in the high-driving-force Tafel regime; the dissolution rate $k_{E,N}$ for a silver atom with $N$ near neighbors was written as $k_{E,N} = \nu_E \exp(-(N\varepsilon - \phi)/k_B T)$, where $\nu_E = 10^4$ $sec^{-1}$ is an attempt frequency determined by the exchange-current density in the BV equation and $\phi$ is the overpotential. For the figure, $\phi$ = 1.75 eV, $\varepsilon/k_B T$ = 5.51.

Figure 3. Comparison of experimental and simulated current-potential behavior. (A) Current-potential behavior for varying Ag-Au alloy compositions (at % Au) dealloyed in 0.1 $M$ $HClO_4$ + 0.1 $M$ $Ag^+$ (reference electrode 0.1 $M$ $Ag^+/Ag$). (B) Simulated current-potential behavior of Ag-Au alloys. (C) Comparison of experimental (line) and simulated (triangles) critical potentials; the zero of overpotential has been set equal to the onset of dissolution of pure silver both in simulation and in experiment.

Figure 4. Simulated evolution of an artificial pit in $Au_{10\%}Ag_{90\%}$ (at. %), $\phi$ = 1.8 eV. Cross-sections along the $(11\bar{1})$ plane defined by the green line in (A) are shown below each plan view. (A) The initial condition is a surface fully passivated with gold except within a circular region (the "artificial pit"). (B) After 1 sec., the pit has penetrated a few monolayers into the bulk. Note how there are fewer gold clusters near the sidewall than at the center of the pit. (C) After 10 sec., a gold cluster has nucleated in the center of the pit. (D) At 100 sec., the pit has split into multiple pits; each will continue to propagate into the bulk to form a porous structure like that in Figure 2.

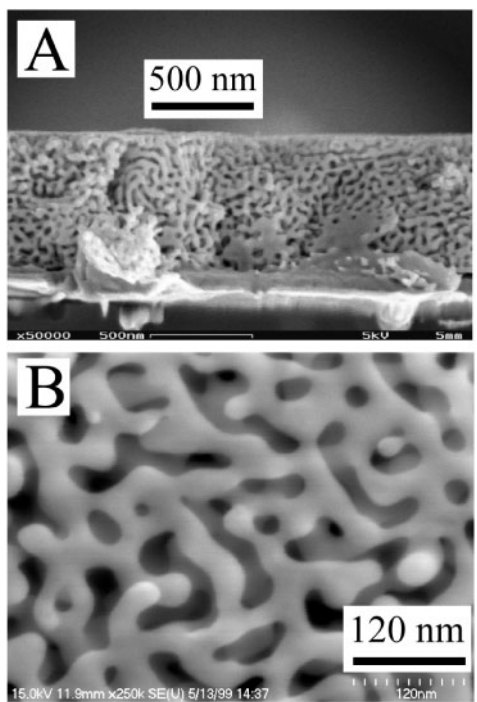

Figure 1.

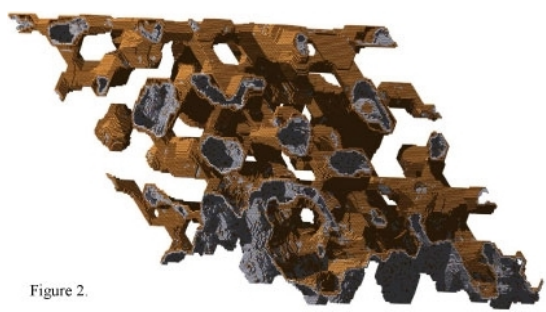

Figure 2.

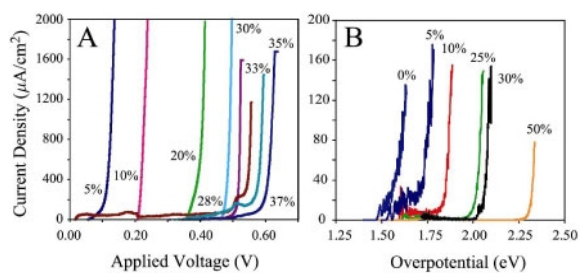

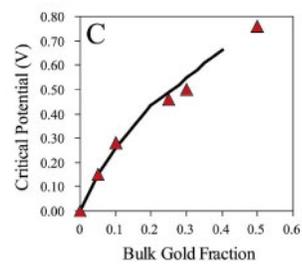

Figure 3.

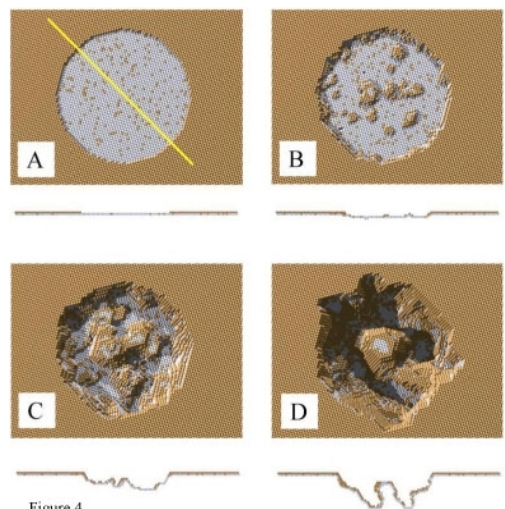

Figure 4.